\documentclass{PoS}

\title{The VINCIA Antenna Shower for Hadron Colliders}

\ShortTitle{The VINCIA Antenna Shower for Hadron Colliders}

\author{\speaker{Peter Skands}\\
        School of Physics and Astronomy, Monash University, VIC-3800, Australia\\
        E-mail: \email{peter.skands@monash.edu}}

\author{Nadine Fischer\\
        School of Physics and Astronomy, Monash University, VIC-3800, Australia
}

\author{Stefan Prestel\\
       SLAC National Accelerator Laboratory, Menlo Park, CA 94025, USA 
}

\author{Mathias Ritzmann\\
        Nikhef, Theory Group, Science Park 105, 1098 XG Amsterdam, The Netherlands
}

\abstract{We summarise the main features of VINCIA's antenna-based
  treatment of QCD initial- and final-state showers, which includes
  iterated tree-level matrix-element corrections and automated evaluations of
  perturbative shower uncertainties. The latter are computed on the
  fly and are cast as a set of alternative weights for each generated
  event. The resulting algorithm has been made publicly available as a
plug-in to the PYTHIA 8 event generator.} 

\FullConference{38th International Conference on High Energy Physics\\
		3-10 August 2016\\
		Chicago, USA}

\begin{document}

\section{Introduction}

Monte Carlo event generators~\cite{Buckley:2011ms,Skands:2012ts} 
provide computer-simulated ``events'',  analogous to those 
produced in real-world high-energy particle collisions. As
such, MC generators represent the most
detailed --- if not always the most accurate --- theoretical reference
calculations in particle physics, and the play crucial roles in a wide
range of contexts, from explicit tests of competing models of
dynamical phenomena, to evaluating the
effects of various types of corrections or cuts and  gauging the sensitivity of proposed
physical observables to salient model parameters. 

From an author perspective, the
\emph{fidelity} of the physics modelling hinges on two main aspects: the
reliability of the approximations made in the various stages of the
calculation/simulation, and the extent to which all relevant physical
phenomena are accounted for. To some extent, these are of course two
sides of the same coin: which effects may be assumed to be
negligible, and what level of precision is achieved for
those that are not?  To complicate matters, the answer to both of these
questions tends to depend on
which observable is being computed, something which 
the MC generator itself cannot control; the user may choose to 
evaluate any (set of) observables on the generated events, from highly 
inclusive total cross sections, to the most exclusive (even
infrared-sensitive) hadron-level observables. A
``good'' event generator is expected to provide a ``reasonable approximation'' 
for the full range of possible observables. 

General-purpose event generators, like PYTHIA
\cite{Sjostrand:2014zea}, HERWIG~\cite{Bellm:2015jjp}, and
SHERPA~\cite{Gleisberg:2008ta}, employ an ordered, top-down 
approach, starting with the
hardest (short-distance), perturbative physics, and ending 
with the softest (long-distance) effects of hadronisation and hadron
decays. This ordering reflects that the long-distance physics is ``seeded'' by
the short-distance physics, from which it also factorises in the limit of
infinite scale separation. Following a hard, perturbative process, MC
generators add the effects of:
\begin{itemize}
\item Initial- and final-state radiation (ISR \& FSR showers);
\item (Sequential) resonance decays (decays of top quarks, Z/W/H
  bosons, and BSM particles);
\item Soft physics: underlying event, hadronisation, hadron and $\tau$
  decays, beam remnant fragmentation, 
  colour reconnections, Bose-Einstein correlations, etc.
\end{itemize}
In this work, we review the recent development of a new formalism for
ISR \& FSR showers, implemented in the publicly available VINCIA
code~\cite{Fischer:2016vfv}, which is written as a plug-in to the PYTHIA 8 event
generator~\cite{Sjostrand:2014zea}. 

\section{VINCIA's Antenna Showers}

Whereas traditional parton showers, such as those implemented in the
standalone version of PYTHIA~\cite{Sjostrand:2004ef,Corke:2010yf}, are
based on an iterated sequence of
 $1\to 2$ splittings, supplemented to include 4-momentum conservation,
 running couplings, and (approximately) the effects of QCD coherence,
 VINCIA is based on the QCD antenna
formalism~\cite{Azimov:1986sf,Gustafson:1987rq,Kosower:1997zr,GehrmannDeRidder:2005cm,Giele:2007di},
which in the MC shower context was originally pioneered by the ARIADNE
program~\cite{Gustafson:1987rq,Lonnblad:1992tz}. In this formalism,
splittings are regarded as occurring inside
colour-connected \emph{pairs} of partons called ``antennae'' (or ``Lund
dipoles''). The associated $2\to 3$ antenna radiation functions
reduce to the standard DGLAP kernels in the collinear
limits, but importantly they also reproduce the soft eikonal factors
in the limit of soft-gluon emission, which assures coherence at the
leading-colour level independently of the choice of
evolution variable and without having to impose any additional vetos
on emission angles. The corresponding $2\to 3$ antenna 
phase-space mappings represent exact, on-shell factorisations of the
$(n+1)/n$-parton phase spaces. They are Lorentz invariant, conserve
4-momentum, and are valid over all of phase space, not just in the
limits. Relative to ARIADNE, the new
elements introduced in VINCIA are:
\begin{itemize}
\item Iterated matrix-element corrections (MECs), based on and
  extending the formalism developed
  in~\cite{Sjostrand:1985xi,Norrbin:2000uu,Giele:2011cb}. In the
  current version of the VINCIA code, these 
  corrections are included up to ${\cal O}(\alpha_s^3)$ for $pp\to Z/W/H +
  \mathrm{jets}$ and up to ${\cal O}(\alpha_s^4)$ for $Z/W/H\to
  \mathrm{jets}$ and for $pp \to \mathrm{jets}$, with matrix elements
  generated by MADGRAPH~\cite{Alwall:2007st,Alwall:2011uj}.  
\item Backwards antenna evolution for ISR, based on and extending the
  formalism developed in
  \cite{Sjostrand:1985xi,Winter:2007ye,Ritzmann:2012ca}. In the
  current implementation, branchings in initial-initial antennae
  (i.e., where both of the parent partons are incoming) do impart
a transverse recoil to the hard system (essential, e.g., to build up
the transverse-momentum spectrum of the lepton pair in $pp \to Z/\gamma^* \to
\ell^+\ell^-$), while branchings in
initial-final antennae do not. We regard the latter as a shortcoming,
necessary to obtain tractable expressions for this first version of
VINCIA for hadron colliders, and to be improved upon in future work. 
\item Automated shower uncertainty estimates, formulated as a
  set of alternative weights for each event, which are calculated on
  the fly, representing variations of renormalisation scales,
  non-singular terms in the antenna functions, colour factors,
  etc. This is based on the formalism proposed by us in \cite{Giele:2011cb},
  which was recently extended also to standalone
  PYTHIA~\cite{Mrenna:2016sih} (accompanied by an all-orders proof of
  the method), with similar implementations having
  appeared in HERWIG~\cite{Bellm:2016voq} and SHERPA~\cite{Bothmann:2016nao}. 
\item Smooth ordering; based on the proposal by us~\cite{Giele:2011cb}
  to allow branching sequences that are not strongly ordered in the
  evolution variable, suppressed by an additional power of $1/Q^2$. 
\end{itemize}
The choice of evolution variable, $Q$, is taken to be inversely proportional
to the leading singularities of the relevant antenna functions;
$p_\perp$ for gluon emissions, and virtuality for processes that
only contain a single pole (such as $g\to q\bar{q}$ splittings). In
\cite{Hartgring:2013jma}, the former was shown to correctly reproduce
the logarithms of the virtual ${\cal  O}(\alpha_s^2)$ antenna functions for
$q\bar{q}\to qg\bar{q}$.  

\begin{figure}[t]
\centering\includegraphics*[scale=0.42]{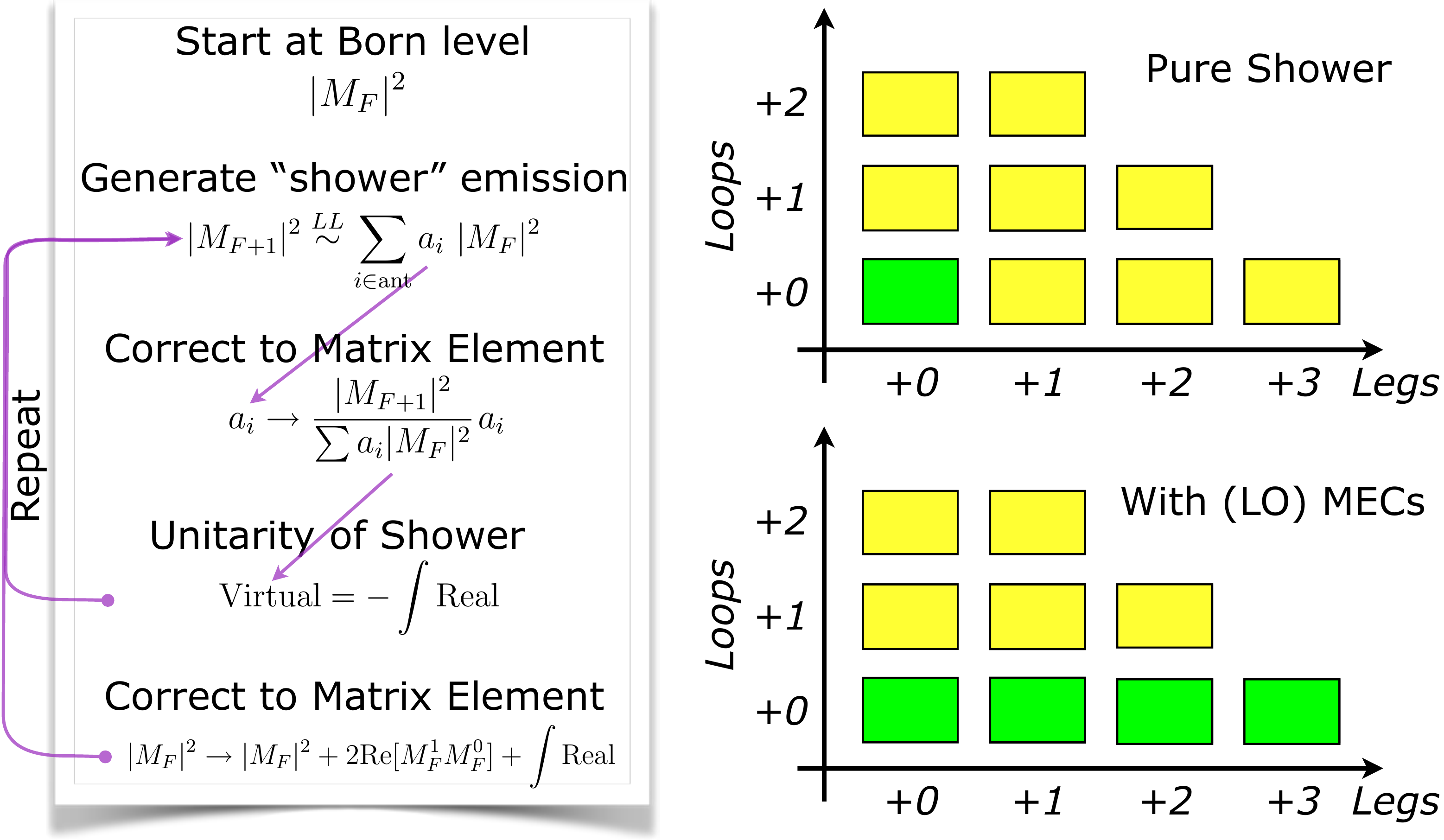}
\caption{Illustration of the iterative application of matrix-element
  corrections (MECs) in VINCIA's shower evolution. In the right-hand
  column, yellow boxes represent coefficients computed at
  leading-logarithmic (LL) accuracy,
while green boxes represent the full perturbative matrix element at the
given order. \label{fig:Markov}}
\end{figure}
The iterated application of matrix-element corrections (MECs) is
illustrated schematically in fig.~\ref{fig:Markov}. Starting from a
Born-level matrix element (squared) for an arbitrary final state, $F$,  the
antenna shower is used to generate an LL approximation to the matrix
element for $F+1$ partons, 
\begin{equation}
|M_{F+1}|^2 \stackrel{LL}{\sim} \sum_{i\in \mathrm{ant}} a_i |M_F|^2~,
\end{equation}
with $a_i$ the appropriate antenna functions, including coupling and
colour factors. Due to the clean antenna phase-space factorisation, no
Jacobian factors appear in this expression, and all the momenta
appearing in both $M_F$ and $M_{F+1}$ are on shell. The probability to
accept the branching is then reweighted by the MEC factor,
\begin{equation}
P_{\mathrm{MEC}} = \frac{|M_{F+1}|^2}{\sum a_i |M_F|^2}~,
\end{equation}
so that the sum of the corrected  branching probabilities,
$P_{\mathrm{MEC}}\times a_i$ reproduces the full (LO) matrix element,
$|M_{F+1}|^2$. Again due to the clean phase-space factorisation, and
introducing a Markovian definition of the evolution
variable~\cite{Giele:2011cb,Fischer:2016vfv}, this procedure can be
iterated for the second and subsequent branchings, until one ``runs out''
of (LO) matrix elements. As the left-hand column indicates, analogous
corrections can also be defined at the loop level, employing
unitarity, though this has so far only been done for the specific case
of $Z \to 3$ jets~\cite{Hartgring:2013jma}. For the first emission,
this formalism reduces to that of PYTHIA for the tree-level
correction~\cite{Sjostrand:1985xi,Norrbin:2000uu}, and to that of
POWHEG for the loop-level correction~\cite{Nason:2004rx,Frixione:2007vw}. What 
is unique to VINCIA's approach is that the corrections can be iterated
for multiple emissions, as demonstrated in~\cite{Giele:2011cb,Hartgring:2013jma,Fischer:2016vfv}. 

For further details on VINCIA's shower, matrix-element corrections,
and automated uncertainty variations, see ~\cite{Fischer:2016vfv} and
references therein. The code can be downloaded from 
\begin{center}
\texttt{http://vincia.hepforge.org}
\end{center}
where an extensive online HTML User Reference is also available. 

\paragraph{Acknowledgements:} We thank the convenors of the
ICHEP strong-interactions session, in particular E.~Nurse, for the
invitation to present our work. SP is supported by the US Department
of Energy under contract DE-AC02-76SF00515. The work of MR was
supported by the European Research Council under Advanced Investigator
Grant ERC-AdG-228301 and ERC Advanced Grant no.~320651, ``HEP-
GAME''. PS is the recipient of an Australian Research Council Future
Fellowship, FT130100744: ``Virtual Colliders: high-accuracy models for
high energy physics''.

\bibliographystyle{utphys}
\bibliography{ants}

\providecommand{\href}[2]{#2}\begingroup\raggedright\begin{thebibliography}{10}

\bibitem{Buckley:2011ms}
A.~Buckley {\em et~al.}, ``{General-purpose event generators for LHC
  physics},'' \href{http://dx.doi.org/10.1016/j.physrep.2011.03.005}{{\em Phys.
  Rept.} {\bfseries 504} (2011) 145--233},
\href{http://arxiv.org/abs/1101.2599}{{\ttfamily arXiv:1101.2599 [hep-ph]}}.

\bibitem{Skands:2012ts}
P.~Skands, \href{http://dx.doi.org/10.1142/9789814525220_0008}{``{Introduction
  to QCD},''} in {\em {Proceedings, Theoretical Advanced Study Institute in
  Elementary Particle Physics: Searching for New Physics at Small and Large
  Scales (TASI 2012): Boulder, Colorado, June 4-29, 2012}}, pp.~341--420.
\newblock 2013.
\newblock
\href{http://arxiv.org/abs/1207.2389}{{\ttfamily arXiv:1207.2389 [hep-ph]}}.
\newblock

\bibitem{Sjostrand:2014zea}
T.~Sj{\"o}strand, S.~Ask, J.~R. Christiansen, R.~Corke, N.~Desai, P.~Ilten,
  S.~Mrenna, S.~Prestel, C.~O. Rasmussen, and P.~Z. Skands, ``{An Introduction
  to PYTHIA 8.2},'' \href{http://dx.doi.org/10.1016/j.cpc.2015.01.024}{{\em
  Comput. Phys. Commun.} {\bfseries 191} (2015) 159--177},
\href{http://arxiv.org/abs/1410.3012}{{\ttfamily arXiv:1410.3012 [hep-ph]}}.

\bibitem{Bellm:2015jjp}
J.~Bellm {\em et~al.}, ``{Herwig 7.0/Herwig++ 3.0 release note},''
  \href{http://dx.doi.org/10.1140/epjc/s10052-016-4018-8}{{\em Eur. Phys. J.}
  {\bfseries C76} no.~4, (2016) 196},
\href{http://arxiv.org/abs/1512.01178}{{\ttfamily arXiv:1512.01178 [hep-ph]}}.

\bibitem{Gleisberg:2008ta}
T.~Gleisberg, S.~Hoeche, F.~Krauss, M.~Schonherr, S.~Schumann, F.~Siegert, and
  J.~Winter, ``{Event generation with SHERPA 1.1},''
  \href{http://dx.doi.org/10.1088/1126-6708/2009/02/007}{{\em JHEP} {\bfseries
  02} (2009) 007},
\href{http://arxiv.org/abs/0811.4622}{{\ttfamily arXiv:0811.4622 [hep-ph]}}.

\bibitem{Fischer:2016vfv}
N.~Fischer, S.~Prestel, M.~Ritzmann, and P.~Skands, ``{Vincia for Hadron
  Colliders},''
\href{http://arxiv.org/abs/1605.06142}{{\ttfamily arXiv:1605.06142 [hep-ph]}}.

\bibitem{Sjostrand:2004ef}
T.~Sj{\"o}strand and P.~Z. Skands, ``{Transverse-momentum-ordered showers and
  interleaved multiple interactions},''
  \href{http://dx.doi.org/10.1140/epjc/s2004-02084-y}{{\em Eur. Phys. J. C}
  {\bfseries 39} (2005) 129},
\href{http://arxiv.org/abs/hep-ph/0408302}{{\ttfamily arXiv:hep-ph/0408302
  [hep-ph]}}.

\bibitem{Corke:2010yf}
R.~Corke and T.~Sj{\"o}strand, ``{Interleaved Parton Showers and Tuning
  Prospects},'' \href{http://dx.doi.org/10.1007/JHEP03(2011)032}{{\em JHEP}
  {\bfseries 03} (2011) 032},
\href{http://arxiv.org/abs/1011.1759}{{\ttfamily arXiv:1011.1759 [hep-ph]}}.

\bibitem{Azimov:1986sf}
Y.~I. Azimov, Y.~L. Dokshitzer, V.~A. Khoze, and S.~I. Troian, ``{The String
  Effect and QCD Coherence},''
\href{http://dx.doi.org/10.1016/0370-2693(85)90709-9}{{\em Phys. Lett. B}
  {\bfseries 165} (1985) 147--150}.

\bibitem{Gustafson:1987rq}
G.~Gustafson and U.~Pettersson, ``{Dipole Formulation of QCD Cascades},''
\href{http://dx.doi.org/10.1016/0550-3213(88)90441-5}{{\em Nucl. Phys. B}
  {\bfseries 306} (1988) 746}.

\bibitem{Kosower:1997zr}
D.~A. Kosower, ``{Antenna factorization of gauge theory amplitudes},''
  \href{http://dx.doi.org/10.1103/PhysRevD.57.5410}{{\em Phys. Rev. D}
  {\bfseries 57} (1998) 5410--5416},
\href{http://arxiv.org/abs/hep-ph/9710213}{{\ttfamily arXiv:hep-ph/9710213
  [hep-ph]}}.

\bibitem{GehrmannDeRidder:2005cm}
A.~Gehrmann-De~Ridder, T.~Gehrmann, and E.~W.~N. Glover, ``{Antenna subtraction
  at NNLO},'' \href{http://dx.doi.org/10.1088/1126-6708/2005/09/056}{{\em JHEP}
  {\bfseries 09} (2005) 056},
\href{http://arxiv.org/abs/hep-ph/0505111}{{\ttfamily arXiv:hep-ph/0505111
  [hep-ph]}}.

\bibitem{Giele:2007di}
W.~T. Giele, D.~A. Kosower, and P.~Z. Skands, ``{A simple shower and matching
  algorithm},'' \href{http://dx.doi.org/10.1103/PhysRevD.78.014026}{{\em Phys.
  Rev. D} {\bfseries 78} (2008) 014026},
\href{http://arxiv.org/abs/0707.3652}{{\ttfamily arXiv:0707.3652 [hep-ph]}}.

\bibitem{Lonnblad:1992tz}
L.~L{\"o}nnblad, ``{ARIADNE version 4: A Program for simulation of QCD cascades
  implementing the color dipole model},''
\href{http://dx.doi.org/10.1016/0010-4655(92)90068-A}{{\em Comput. Phys.
  Commun.} {\bfseries 71} (1992) 15--31}.

\bibitem{Sjostrand:1985xi}
T.~Sj{\"o}strand, ``{A Model for Initial State Parton Showers},''
\href{http://dx.doi.org/10.1016/0370-2693(85)90674-4}{{\em Phys. Lett. B}
  {\bfseries 157} (1985) 321}.

\bibitem{Norrbin:2000uu}
E.~Norrbin and T.~Sj{\"o}strand, ``{QCD radiation off heavy particles},''
  \href{http://dx.doi.org/10.1016/S0550-3213(01)00099-2}{{\em Nucl. Phys. B}
  {\bfseries 603} (2001) 297--342},
\href{http://arxiv.org/abs/hep-ph/0010012}{{\ttfamily arXiv:hep-ph/0010012
  [hep-ph]}}.

\bibitem{Giele:2011cb}
W.~T. Giele, D.~A. Kosower, and P.~Z. Skands, ``{Higher-Order Corrections to
  Timelike Jets},'' \href{http://dx.doi.org/10.1103/PhysRevD.84.054003}{{\em
  Phys. Rev. D} {\bfseries 84} (2011) 054003},
\href{http://arxiv.org/abs/1102.2126}{{\ttfamily arXiv:1102.2126 [hep-ph]}}.

\bibitem{Alwall:2007st}
J.~Alwall, P.~Demin, S.~de~Visscher, R.~Frederix, M.~Herquet, F.~Maltoni,
  T.~Plehn, D.~L. Rainwater, and T.~Stelzer, ``{MadGraph/MadEvent v4: The New
  Web Generation},''
  \href{http://dx.doi.org/10.1088/1126-6708/2007/09/028}{{\em JHEP} {\bfseries
  09} (2007) 028},
\href{http://arxiv.org/abs/0706.2334}{{\ttfamily arXiv:0706.2334 [hep-ph]}}.

\bibitem{Alwall:2011uj}
J.~Alwall, M.~Herquet, F.~Maltoni, O.~Mattelaer, and T.~Stelzer, ``{MadGraph 5
  : Going Beyond},'' \href{http://dx.doi.org/10.1007/JHEP06(2011)128}{{\em
  JHEP} {\bfseries 06} (2011) 128},
\href{http://arxiv.org/abs/1106.0522}{{\ttfamily arXiv:1106.0522 [hep-ph]}}.

\bibitem{Winter:2007ye}
J.-C. Winter and F.~Krauss, ``{Initial-state showering based on colour dipoles
  connected to incoming parton lines},''
  \href{http://dx.doi.org/10.1088/1126-6708/2008/07/040}{{\em JHEP} {\bfseries
  07} (2008) 040},
\href{http://arxiv.org/abs/0712.3913}{{\ttfamily arXiv:0712.3913 [hep-ph]}}.

\bibitem{Ritzmann:2012ca}
M.~Ritzmann, D.~A. Kosower, and P.~Skands, ``{Antenna Showers with Hadronic
  Initial States},''
  \href{http://dx.doi.org/10.1016/j.physletb.2012.12.003}{{\em Phys. Lett. B}
  {\bfseries 718} (2013) 1345--1350},
\href{http://arxiv.org/abs/1210.6345}{{\ttfamily arXiv:1210.6345 [hep-ph]}}.

\bibitem{Mrenna:2016sih}
S.~Mrenna and P.~Skands, ``{Automated Parton-Shower Variations in Pythia 8},''
  {\em Submitted to: Phys. Rev. D} (2016) ,
\href{http://arxiv.org/abs/1605.08352}{{\ttfamily arXiv:1605.08352 [hep-ph]}}.

\bibitem{Bellm:2016voq}
J.~Bellm, S.~Pl{\"a}tzer, P.~Richardson, A.~Si{\'o}dmok, and S.~Webster,
  ``{Reweighting Parton Showers},''
  \href{http://dx.doi.org/10.1103/PhysRevD.94.034028}{{\em Phys. Rev.}
  {\bfseries D94} no.~3, (2016) 034028},
\href{http://arxiv.org/abs/1605.08256}{{\ttfamily arXiv:1605.08256 [hep-ph]}}.

\bibitem{Bothmann:2016nao}
E.~Bothmann, M.~Sch{\"o}nherr, and S.~Schumann, ``{Reweighting QCD
  matrix-element and parton-shower calculations},''
\href{http://arxiv.org/abs/1606.08753}{{\ttfamily arXiv:1606.08753 [hep-ph]}}.

\bibitem{Hartgring:2013jma}
L.~Hartgring, E.~Laenen, and P.~Skands, ``{Antenna Showers with One-Loop Matrix
  Elements},'' \href{http://dx.doi.org/10.1007/JHEP10(2013)127}{{\em JHEP}
  {\bfseries 10} (2013) 127},
\href{http://arxiv.org/abs/1303.4974}{{\ttfamily arXiv:1303.4974 [hep-ph]}}.

\bibitem{Nason:2004rx}
P.~Nason, ``{A New method for combining NLO QCD with shower Monte Carlo
  algorithms},'' \href{http://dx.doi.org/10.1088/1126-6708/2004/11/040}{{\em
  JHEP} {\bfseries 11} (2004) 040},
\href{http://arxiv.org/abs/hep-ph/0409146}{{\ttfamily arXiv:hep-ph/0409146
  [hep-ph]}}.

\bibitem{Frixione:2007vw}
S.~Frixione, P.~Nason, and C.~Oleari, ``{Matching NLO QCD computations with
  Parton Shower simulations: the POWHEG method},''
  \href{http://dx.doi.org/10.1088/1126-6708/2007/11/070}{{\em JHEP} {\bfseries
  11} (2007) 070},
\href{http://arxiv.org/abs/0709.2092}{{\ttfamily arXiv:0709.2092 [hep-ph]}}.

\end{thebibliography}\endgroup

\end{document}